\documentclass{Rinton-P9x6}

\begin{document}

\title{Early Universe cosmology and supergravity}

\author{Anupam Mazumdar}

\address{McGill University, 3600 University Road, Montreal, Canada H3A~2T8
}

\maketitle

\abstracts{In this talk I will describe the role of supersymmetric 
flat directions in the early Universe cosmology. Particularly I will 
discuss how supergravity effects lift the flat direction potential and 
leads to interesting cosmological consequences.}

\section{Supergravity $\&$ cosmology}

Like particle physics cosmology also confronts hierarchical scales;
e.g. Planck scale $M_{p}=2.436\times 10^{18}$~GeV, and big bang 
nucleosynthesis scale $T_{BBN}\sim 1$~MeV, and uncertainties. 
Unfortunately there is no direct evidence of thermal history of 
the Universe beyond $T\sim 1$~MeV. Though there are indirect evidences
which when tied up with particle physics makes a strong case that 
thermal history could go way beyond the nucleosynthesis scale to the
grand unification scale $T\leq \alpha^2M_{p}\sim 10^{16}$~GeV, where
$\alpha\sim 10^{-2}$ is the gauge coupling constant. Note that beyond 
the grand unification scale it might be hard to keep the Universe in 
a thermal equilibrium due to lack of understanding of interactions and 
strong dynamics which could prevail. In this respect supergravity (SUGRA) 
\cite{nilles84} plays an important role in constraining thermal history 
of the early Universe, see~\cite{Enqvist}.

The superpartner of graviton is a spin $3/2$ gravitino with helicity 
states $\pm 3/2,~\pm 1/2$. The latter one is mainly the goldstino 
mode which is eaten by the super-Higgs mechanism~\cite{deser77} 
when supersymmetry is spontaneously broken. Typically ${\cal N}=1$
supersymmetry (SUSY) is broken in a hidden sector by some 
non-perturbative dynamics, e.g. gaugino condensation~\cite{nilles84}. 
SUSY breaking  is then
mediated via gravitational (or possibly other) couplings to the observable
sector in such a way that sfermions and gauginos get masses of order
electroweak scale \cite{nath84,nilles84}. In addition, the gravitino also
gets a mass, which in the simplest gravity mediated models is of 
order ${\cal O}(100)$~GeV.

Although gravitino interactions with matter are suppressed by the Planck
mass, they can be generated in great abundances very close to the Planck
scale \cite{weinberg82}. Inflation would dilute their number density, 
but during reheating they would be regenerated
though scattering of gauge and gaugino quanta, with adverse consequences
\cite{ellis84b}. Thermal regeneration of gravitino abundance is harmless
for BBN (gravitino decay products can enhance the abundance of 
$D+^{3}He$ due to photo fission of $^{4}He$), provided the reheat 
temperature of the Universe is $T_{rh} \leq (10^7-10^{10}) $~GeV, 
for a range of gravitino mass $100~{\rm GeV}\leq m_{3/2}\leq 10~{\rm TeV}$
\cite{ellis85b,sarkar96}. Gravitinos could also be produced by 
non-perturbative processes during inflaton oscillations~\cite{maroto99}. 
Note that expanding Universe breaks supersymmetry and normally the scale 
is set by the Hubble expansion, which could be larger than the hidden sector 
supersymmetry breaking.

It is this early Universe supersymmetry breaking scale plays interesting
role. SUGRA plays a very important role 
in higher dimensions motivated by superstring theory. In this respect 
cosmological evolution of dilaton and moduli is an important issue 
which has to be dealt separately. However recently dynamics of dilaton
and moduli has been addressed in presence of Ramond Ramond fluxes~\cite{Frey}.
Inflationary scenarios~\cite{Liddle} have also been proposed where 
${\cal N}=1$, $11$ dimensional SUGRA model is squashed on a 
seven sphere. The resulting potentials turn out to be some combination 
of multi-field exponential potentials of type 
$c_{1}\exp{\sum \alpha_{i}\phi_{i}}+c_{2}\exp{\sum\beta_{i}\psi_{i}}+..$.

In this talk we will not concentrate on finding inflationary solutions, but
we will describe the dynamics of the flat directions within minimal 
supersymmetric standard model (MSSM). Flat directions is a property of
any SUSY theory and here we will argue how interesting cosmology 
arises when we take into account of SUGRA effects.


\section{Status of outstanding issues in cosmology}

Before we discuss flat directions let us give a brief overview of 
the present status of cosmological issues. The recent W-MAP data 
has provided the precision measurement of cosmological parameters,
e.g. Hubble expansion, fluctuations in the temperature, perturbation
spectrum and its slope, running spectral index, etc~\cite{WMAP}, 
all suggest that perhaps an era of inflation might have occurred. 
Despite these observational successes, inflaton sector is still eluding. 
There are many plausible models which can provide more or less correct 
features but none of them are well motivated from particle physics. The main
reason is that inflaton is always treated as a SM gauge singlet. 
This leads to a wild speculation of inflaton coupling to the
SM fields and the scale of inflation ranging from 
$10^{13}\leq H\leq 10^{3}$~GeV. Extremely low scale inflation
model is possible to construct if the string scale is as low as 
TeV, e.g., \cite{anu}, for a TeV scale inflation number of e-foldings
required for galaxy formation is as low as $43$.

Baryogenesis is an indirect evidence of cosmology and particle physics
beyond the Higgs scale, because baryon asymmetry of order one part in 
$10^{10}$ is sought for the synthesis of light elements at $T\sim 1$~MeV. 
Within SM it is possible to satisfy all three criteria for baryogenesis,
e.g. baryon number violating processes (due to accidental $B+L$ violation
arising from the SM sphalerons), $C$ and $CP$ violation, and out of 
equilibrium condition. Inspite of all the necessary ingredients within SM
baryogenesis is ruled out because there is not enough strong first order
phase transition (success of baryogenesis requires Higgs mass to be
$m_{H}\leq 72$~GeV~\cite{rubakov}, which is already ruled out by the 
LEP data, see~\cite{pdg}). 
Within MSSM there is a slim hope because the parameter space for Higgs is
$110~{\rm GeV}\leq m_{H}\leq 115~{\rm GeV}$, and the lightest stop 
to be lighter than the top quark mass
$105~{\rm GeV}\leq m_{\tilde t_{R}}\leq 165~{\rm GeV}$~\cite{carlos}.

There are other baryogenesis schemes such as thermal leptogenesis in
models where right handed neutrino mass appears naturally~\cite{Buchmuller},
or non-thermal leptogenesis~\cite{Dutta}.
However thermal leptogenesis within $SO(10)$ requires generally large
reheat temperature~$T_{rh}\geq 10^{10}$, which is generically considered 
to be tight given the gravitino overproduction constraints. One might resort
for non-thermal leptogenesis scenarios, but unfortunately they bear the 
uncertainties of reheat temperature of the Universe. Though they could be
tested from CMB experiments~\cite{maz}. In this respect
MSSM flat directions can act as a savior~\cite{AD,dine96}, because a generic 
flat direction carries global $U(1)$ baryon and/or lepton number which 
can be broken due to soft A-terms, which we will discuss briefly, 
for details see~\cite{Enqvist}.

Other important issues are the cold dark matter generation which comprise 
$30\%$ of the total energy density, and the cosmological constant which 
has the largest share of $70\%$ at the present moment. The cold dark 
matter is well served by supersymmetric candidate neutralino (We have 
had many talks on this issue in this conference, see \cite{olive}). 
The cosmological 
constant is a real issue for which we have NO understanding. The major
problem is why the cosmological constant is so small, and why it is 
dominating now. There are some resolution of the smallness and why now 
for cosmological constant in terms of non-perturbative $\theta$-vacuum, 
see~\cite{Jaikumar}.

\section{A brief review of MSSM flat directions}\label{subsec:prod}

At the level of renormalizable terms, SUSY field theories
generically have infinitely degenerate vacua. This is a consequence of
SUSY and gauge symmetries (and discrete symmetries such as
$R$-parity) of the Lagrangian. Therefore, in general there are a number of
directions in the space of scalar fields, collectively called the moduli
space, where the scalar potential is identically zero. In low
energy SUSY theories such classical degeneracy is accidental
and is protected from perturbative quantum corrections by a
non-renormalization theorem \cite{gwr79}. In principle the
degeneracies could be lifted by non-perturbative effects. However such
effects are likely to be suppressed exponentially and thus unimportant
because all the couplings of low energy theories are typically weak even
at relatively large vevs. Therefore in a SUSY limit when
$M_{\rm p}\rightarrow \infty$, the potential for the flat direction
always vanishes.

However there is an effective potential for the flat direction arises 
as a result of SUSY breaking terms and higher dimensional 
operators in the superpotential. In this sense MSSM flat directions 
are only approximately flat at vevs larger than the SUSY
breaking scale.

The SUSY scalar potential $V$ is the sum of the F- and D-terms and reads
\begin{equation}
V= \sum_i |F_i|^2+\frac 12 \sum_a g_a^2D^aD^a\,,~~
F_i\equiv {\partial W_{MSSM}\over \partial \phi_i},~~D^a=\phi^{i~\dagger} T^a
\phi_{i}~.
\label{fddefs}
\end{equation}
Here we have assumed that $\phi_i$ transforms under a gauge group
$G$ with the generators of the Lie algebra given by $T^{a}$.

An example of a D-and F-flat direction is provided by $LH_{u}$.
\begin{equation}
\label{example}
H_u=\frac1{\sqrt{2}}\left(\begin{array}{l}0\\ \phi\end{array}\right),~
L=\frac1{\sqrt{2}}\left(\begin{array}{l}\phi\\ 0\end{array}\right)~,
\end{equation}
We can denote $\phi$ as a complex field parameterizing the flat direction,
or the order parameter, or the AD field. All the other fields are
set to zero. In terms of the composite gauge invariant operators,
we would write $X_m=LH_{u}~(m=2)$.

From Eq.~\ref{example} one clearly obtains $
F_{H_u}^*=\lambda_uQ\bar u +\mu H_d=F_{L}^*=\lambda_dH_d\bar e\equiv 0$
for all $\phi$. However there exists a non-zero F-component given
by $F^*_{H_d}=\mu H_u$. Since $\mu$ can not be much larger than the
electroweak scale $M_W\sim {\cal O}(1)$~TeV, this contribution is of
the same order as the soft supersymmetry breaking masses, which are
going to lift the degeneracy. Therefore, following \cite{dine96}, one may
nevertheless consider $LH_u$ to correspond to a F-flat direction.

The relevant D-terms read
$D^a_{SU(2)}=H_u^\dagger\tau_3H_u+L^\dagger\tau_3L=\frac12\vert\phi\vert^2
-\frac12\vert\phi\vert^2\equiv 0$. Therefore $LH_u$ direction is also D-flat. 
All the MSSM flat directions are listed in~\cite{gherghetta} following
\cite{buccella}.

\section{Lifting flat directions}

Vacuum degeneracy along a flat direction can be broken in two ways:
by supersymmetry breaking, or by higher order non-renormalizable
operators appearing in the effective low energy theory.

A particular class of non-renormalizable interaction terms induced by the
inflaton arise for a K\"ahler potential which has a form
\cite{gaillard95,dine96,bagger95}, and which generates terms like
\begin{equation}
\label{Iphicoupl}
K =\int d^4\theta \frac{1}{M_{\rm P}^2}(I^{\dagger}I)(\phi^{\dagger}\phi)\,,
~~{\cal L}= \frac{\rho_{I}}{M_{\rm P}^2}\phi^{\dagger}\phi =3H_I^2
\phi^{\dagger}\phi \,,
\end{equation}
where $I$ is the inflaton whose energy density
$\rho \approx \langle \int d^4\theta I^{\dagger}I\rangle$ dominates
during inflation, and $\phi$ is the flat direction. In the above  
$H_I$ is the Hubble parameter during inflation. In addition, there are
also inflaton-induced supergravity corrections
to the flat direction. By inspecting ${\cal N}=1$ supergravity potential,
one finds the following terms like
\begin{equation}
\label{mflat}
V(\phi)=H^2M_{\rm P}^2 f\left(\frac{\phi}{M_{\rm P}}\right)\,,~~
m^2_{\phi}=\left(2+\frac{F^{\ast}_{I}F_{I}}{V_{F}(I)+V_{D}(I)}\right)H^2\,.
\end{equation}
where $f$ is some function, and the effective mass for inflaton 
is written for a minimal choice of flat direction K\"ahler potential 
$K(\phi^{\dagger},\phi)=\phi^{\dagger}\phi$. 

In purely D-term inflation~\cite{binetruy} there is no Hubble
induced mass correction to the flat direction during inflation
because $F_{I}=0$. However, when D-term inflation ends, the energy density
stored in the D-term is converted to an F-term and to kinetic energy of
the inflaton. Thus again a mass term $m_{\phi}^2 =\pm {\cal O}(1)H^2$
appears naturally, however the overall sign is undetermined \cite{kolda98}.
For non-flat K\"ahler potential such as in no-scale models
$K\sim \ln (z+z^*+\phi_i^\dagger\phi)$, where $z$ belongs to supersymmetry
breaking sector, and $\phi_{i}$ belongs to the matter sector, and both are
measured in terms of reduced Planck mass (for a review, see~\cite{lahanas87}
the mass term for the flat direction in presence of inflationary potential  
turns out to be $m^2_{\phi}\sim 10^{-2}H_{I}^2$~\cite{gaillard95,campbell99}.

Besides supergravity correction the flat directions can be lifted by the 
non-renormalizable contributions such as~\cite{dine96}
\begin{equation}
W=\frac{h}{d M^{d-3}}\phi^d\,,~~W={h\over M^{d-3}}\psi\phi^{d-1}~\,,
~~{\rm and}~~V(\phi)={\vert\lambda\vert^2\over M^{2d-6}}(\phi^*\phi)^{d-1}\,.
\end{equation}
where the dimensionality of the effective scalar operator is $d$,
and $h$ is a coupling constant which could be complex with
$|h|\sim {\cal O}(1)$. Here $M$ is some large mass, typically of
the order of the Planck mass or the string scale (in the heterotic
case $M \sim M_{GUT}$). 

Vacuum degeneracy will also be lifted by supersymmetry
breaking. It is induced by the soft terms, which in the simplest case read
\begin{equation}
\label{susybreak}
V(\phi)=m_0^2\vert\phi\vert^2+\left[{\lambda A\phi^d\over dM^{d-3}}
+{\rm h.c.}\right]~,
\end{equation}
where SUSY breaking mass $m_0$ and $A$ are typically of the order
of the gravitino mass $m_{3/2}$. An additional soft source for SUSY
breaking are the gaugino masses $m_g$. The $A$-term in Eq.~(\ref{susybreak})
violates the $U(1)$ carried by the flat direction and thus provides the
necessary source for $B-L$ violation in AD baryogenesis. In
general, the coupling $\lambda$ is complex and has an associated  phase
$\theta_\lambda$. Writing
$\phi=\vert\phi\vert \exp(i\theta)$, one obtains a potential
proportional to $\cos(\theta_\lambda+n\theta)$ in the angular
direction. This has $n$ discrete minima for the phase of $\phi$,
at each of which $U(1)$ is broken.

\section{The potential for flat direction}

\subsection{$F$-term inflation}
Let us collect all the terms which contribute to the flat 
direction potential, which in the case of F-term inflation 
can be written as \cite{dine96}
\begin{eqnarray}
\label{adpot0}
V(\phi)&=&-C_{I} H_{I}^2 {|\phi |}^2 + \left(a {\lambda}_d H {{\phi}^d \over d
M^{d-3}} + {\rm h.c.}\right) + m^{2}_{\phi}{|\phi|}^2\nonumber \\
&&+ \left(A_{\phi} {\lambda}_d \frac{{\phi}^d}{dM^{d-3}} + {\rm h.c.}\right)
+|\lambda|^2\frac{|\phi|^{2d-2}}{M^{2d-6}} \,.
\end{eqnarray}
The first and the third terms are the Hubble-induced and low-energy soft
mass terms, respectively, while the second and the fourth terms are the
Hubble-induced and low-energy $A$ terms. The last term is
the contribution from the non-renormalizable superpotential. The coefficients
$|C_{I}|,~a,~\lambda_{d}\sim {\cal O}(1)$, and the coupling 
$\lambda \approx 1/(d-1)!$. Note that low-energy $A_{\phi}$ term is 
dimensionful. 

Note here the importance of the relative sign of the coefficient
$C_{I}$. At large field values the first term dictates the
dynamics of the AD field. If $C_{I}<0$ , the absolute minimum
of the potential is   $\phi =0$ and during inflation the AD field will
settle down to the bottom of the potential roughly in one Hubble time.
In such a case the AD field will not have any interesting classical dynamics.
Its presence will nevertheless be felt because of  quantum
fluctuations. These will be chi-squared in nature since then the classical
energy density of the AD field would be due to its own fluctuations.

If $C_{I} \ll 1$, the AD field takes some time to reach the bottom of
the potential, and if it has a non-zero amplitude after the end of
inflation, its dynamics is  non-trivial.

The most interesting scenario occurs when $C_{I}>0$.  In this
case the absolute value of the AD field settles during inflation to
the minimum given by
\begin{equation}
\label{veveq}
|\phi| \simeq \left({C_{I}\over (d-1) {\lambda}_d} H_{I}M^{d-3}\right)^{1/d-2}
\,.
\end{equation}
Here we have ignored the potential term $\propto a$;
if $C_{I}> 0$, the $a$-term will not change the vev
qualitatively. On the other hand, even for $C_{I}<0$ the potential
Eq.~(\ref{adpot0}) will have a minimum with a non vanishing vev if
$|a|^2 >4(d-1) C_{I}$. However the origin will also be a minimum in this
case. The dynamics then depends on which minimum
the AD field will choose during inflation.

The $a$-term in Eq.~(\ref{adpot0}) violates the global $U(1)$ symmetry
carried by $\phi$. If $|a|$ is ${\cal O}(1)$, the phase $\theta$ of
$\langle \phi \rangle$ is related to the phase of $a$ through $n
\theta + {\theta}_a = \pi$; otherwise $\theta$ will take some random
value, which will generally be of ${\cal O}(1)$. This is the initial
$CP$-violation which is required for baryogenesis/leptogenesis.
In practice, the superpotential term lifting the flat direction
is also the $B$ and $CP$ violating operator responsible for AD baryogenesis,
inducing a baryon asymmetry in the coherently oscillating $\phi$ condensate.

\subsection{D-term inflation}

In D-term inflation one does not get the Hubble induced mass
correction to the flat direction so that $C_{I}=0$.
Also the Hubble induced $a$-term is absent. However the Hubble induced mass
correction eventually dominates once D-term induced inflation comes to an
end. The potential for a generic flat direction during
D-term inflation is given by
\begin{equation}
\label{adpot1}
V(\phi)=m^{2}_{\phi}{|\phi|}^2
+ \left(A_{\phi} {\lambda}_d \frac{{\phi}^d}{dM^{d-3}} + {\rm h.c.}\right)
+|\lambda|^2\frac{|\phi|^{2d-2}}{M^{2d-6}} \,,
\end{equation}
and after the end of inflation the flat direction potential
is given by \cite{kolda98}
\begin{equation}
\label{adpot2}
V(\phi)=\left(m^{2}_{\phi}-C H^2\right){|\phi|}^2
+ \left(A_{\phi} {\lambda}_d \frac{{\phi}^d}{dM^{d-3}} + {\rm h.c.}\right)
+|\lambda|^2\frac{|\phi|^{2d-2}}{M^{2d-6}} \,,
\end{equation}
where $C\sim {\cal O}(1)$. For $C$ positive, the flat direction settles
down to one of its minima given by Eq.~(\ref{veveq}) provided
$\phi\geq \sqrt{m_{\phi}M/\lambda}$, otherwise
\begin{equation}
\label{veveq1}
|\phi| \simeq \left(\frac{2C}{\lambda_{d}A_{\lambda}(d-1)}H(t)^2M^{d-3}
\right)^{1/d-2}\,,
\end{equation}
Note that in this case that the $A$-term is also responsible for
$B$ and/or $L$, and $CP$ violation. Another generic point to remember
is that in $R$-parity conserving models the $B$ and/or $L$ violating operators
must have even dimensions, so that $d=4$ yields  the minimal operator
for AD baryogenesis.

\section{Various applications of MSSM flat directions in cosmology}

Here we briefly recall important application of MSSM flat directions
in the early Universe, and then in the subsequent subsection we will
concentrate on how MSSM flat directions can generate observed density 
fluctuations in the Universe. 

 {\it (1)~Can MSSM flat directions lead to inflation ?}\\
At sufficiently large vevs the non-renormalizable potential 
$\lambda^2 |\phi|^{2d-2}/M^{2d-6}$ dominates over all the 
contributions, provided there is Hubble induced correction. 
Nevertheless the vev of the flat direction never exceeds 
$M_{p}$ required for a successful inflation in 
this case. However one might expect assisted inflation~\cite{Liddle} 
with the help of 300 flat directions within MSSM. Though the 
yield of number of e-foldings is much less than $55$, required 
to make the Universe flat and homogeneous.

{\it (2)~Can MSSM flat direction gives rise to density perturbations ?}\\
For $D$-term inflation MSSM flat directions evolve in the non-renormalizable
potential and feel the quantum fluctuations $\delta \phi\sim H_{I}/2\pi$, 
This is a good news because now we may be able to generate the observed
density perturbations from the MSSM flat direction, instead from the inflaton
sector whose potential we may never be able to construct with full
confidence. Besides the MSSM flat direction potential is more constrained
than some unknown inflaton sectors~\cite{enqvist02}.

{\it (3)~Can MSSM flat direction gives rise to baryogenesis ?}\\
Yes, the flat direction carries global U(1) charge. For $LH_{u}$ it 
is $B-L$. Note that global $U(1)$ is broken only by the $A$-terms.
This provides an ideal condition for $CP$ violation (dynamically),
which leads to helical motion on the field space. As a flat direction
rotates, it generates the asymmetry. The final leptonic asymmetry is 
created when $LH_{u}$ direction completely evaporates. One must  
take into account of thermal corrections: $\propto T^2|\phi|^2$
and $T^4\ln(|\phi^2/M^2|)$. One can further make connection with 
the light neutrino masses and leptogenesis from the sliding vev of 
$LH_{u}$ flat direction. It turns out that the lightest 
left handed neutrino mass to be very small around $10^{-8}$~eV~\cite{fuji}.
Recent analysis shows that it is possible to obtain the right baryon asymmetry
of order one part in $10^{10}$.

{\it (4)~Do MSSM flat directions run ?}\\
One has to consider the running of the soft terms for the 
MSSM flat directions, e.g. $m^2_{\phi}|\phi|^2$ term. Running
of the MSSM flat direction during inflation has been 
considered in Ref.~\cite{Allahverdi}. It was concluded that
$LH_{u}$ flat direction is a likely candidate whose ${\rm mass}^2$
running remain positive during inflation irrespective of the sign of
$C_{I}$. Post-inflationary running can give rise to interesting 
effects depending on low energy supersymmetry breaking schemes,
e.g. gravity, or gauge mediation. The potentials for both gravity 
and gauge mediation can be written, respectively 
as~\cite{enqvist98,enqvist99,Kusenko}
\begin{eqnarray}
\label{pot}
U(\Phi)&=&m_{\phi}^{2}\left(1 +K \log\left(\frac{|\Phi|^{2}}{M^{2}}
\right) \right) |\Phi|^{2}+ \frac{\lambda^{2}|\Phi|^{2(d-1)}
}{M_{\rm P}^{2(d-3)}} + \left( \frac{A_{\lambda}
\lambda \Phi^{d}}{d M_{\rm P}^{d-3}} + h.c.\right)\, \nonumber\\
U(\Phi)&=&m_{\phi}^4\log\left(1+\frac{|\Phi|^2}{m_{\phi}^2}\right)
+\frac{\lambda^2 |\Phi|^{2(d-1)}}{M_{\rm P}^{2(d-3)}}+
\left(\frac{A_{\lambda}\lambda\Phi^{d}}{dM_{\rm P}^{d-3}}+{\rm h.c.}\right)\,.
\end{eqnarray}
where $m_{\phi} \approx 100$~GeV for gravity mediated SUSY breaking,
while $m_{\phi}\sim 1-100$~TeV in the case of gauge mediation. 
$K$ is a parameter which depends on the flat direction, and the 
logarithmic contribution parameterizes the running of the flat 
direction potential with $M=(M_{\rm P}^{d-3}m_{3/2}/|\lambda|)^{(1/d-2)}$. 
In the gravity mediated case $|A_{\lambda}|< dm_{3/2}$, for
gauge mediation $|A_{\lambda}| \leq (10^{-4}-10^{-7})m_{\phi}$,
for $d=4,6$~\cite{jokinen02}. For the squark directions the value of 
$K$ tends to become negative and this has interesting consequences
\cite{McDonald}.

{\it (5)~What is the fate of the flat direction ?}\\
The post inflationary running of the MSSM flat direction plays an important 
role. The flat direction starts oscillating at the SUSY breaking scale 
under the influence of potentials Eqs.~(\ref{pot}) depending on whether
SUSY is broken via gravity mediation or gauge mediation. For a negative
values of $|K|<0$, the leading potential term in gravity mediation case
is $\sim |\phi|^{2-2K}$ shallower than the $\phi^2$. The oscillations
in such a potential on average gives rise to a negative pressure which 
fragments the flat direction~\cite{McDonald}. The linear fluctuations 
grow non-linearly
and form non-topological solitons, known as $B$-balls. The $B$-balls
carry baryon number $\sim 10^{17}-10^{18}$~\cite{enqvist98,enqvist99}, and 
release them via surface 
evaporation, which takes place usually after the electro-weak scale. 
For every baryon number there
are $3$ LSP (which is bino in the case of minimal SUGRA). Therefore MSSM 
flat direction is the only mechanism which can connect baryon number and
cold dark matter abundance within an order of magnitude. It turns out that
binos are over produced and therefore not hailed as a good candidate for
CDM. However it opens the door for Higgsino and/or Wino to be the right 
candidate~\cite{fujihama}. The $Q$-balls are also formed in gauge mediation, 
though in this 
case mass per unit charge of a $Q$ ball can be less than the 
nucleon mass~\cite{kus} (for numerical simulation
see~\cite{Kasuya}). Therefore rendering such $Q$ balls absolutely stable 
under surface evaporation. These $Q$-balls act as a potential candidate for 
CDM with $Q\geq 10^{12}$. Various indirect searches for dark matter 
experiment put constraints on the charge of stable $Q$ balls, which is
right now $Q\geq 10^{22}$~\cite{arafune00}.

The above mentioned applications exhibit how rich MSSM flat directions
are for the early Universe cosmology. This is the most economical approach 
in obtaining a meaningful early Universe in conjunction with supersymmetry.
In the remaining talk I will concentrate upon very new development on
how MSSM flat directions can help generating scale invariant adiabatic 
density fluctuations.


\section{Magic without Magic: generating adiabatic fluctuations}

At the time of last scattering surface the perturbations in photons, 
baryons and CDM can be characterized by an overall curvature perturbation $R$,
and a relative isocurvature component $S$ in the different matter components
\begin{eqnarray}
R=H\frac{\delta\rho}{\dot\rho}\,,~~S=3H\left(\frac{\delta\rho_{\gamma}}
{\dot\rho}-\frac{\delta\rho_{i}}{\dot\rho_{i}}\right)\,.
\end{eqnarray} 
In a pure adiabatic case $S=0$. But in general $R$ and $S$ could be 
correlated. An important point to realize that 
{\it entropy perturbations seed  adiabatic perturbations on large scales}. This
is the key for MSSM flat directions to succeed as a candidate for
generating scale invariant fluctuations. The question arises if there
are roughly $300$ flat directions present during the inflationary epoch 
then what do they do?. As explained earlier inflaton sector need not take
the burden of explaining the density fluctuations. There could be MSSM 
flat directions pioneering this role of generating perturbations.
There are two popular schemes which I briefly sketch in this talk.

{\it (1)~Curvaton scenario:}\\
Quantum fluctuations of the MSSM flat direction during inflation give rise to
isocurvature perturbations. To comply with CMB measurements, these
perturbations have to be converted into adiabatic ones~\cite{sloth,lyth,mar}. 
Such a
conversion takes place when the contribution of the curvaton energy
density $\rho_\phi$ to the total energy density in the universe grows,
i.e., with the increase of
\be
r =\frac{3 \rho_\phi}{4 \rho_\gamma + 3 \rho_\phi}\,.
\label{r}
\ee
Here $\rho_\gamma$ is the energy in the radiation bath from inflaton
decay (a hidden sector inflaton reheats into hidden sector relativistic
degrees of freedom). Non-Gaussianity of the produced adiabatic perturbations
requires the MSSM flat direction to contribute more than 1\% to the energy
density of the universe at the time of decay, that is 
$r_{\rm dec} > 0.01$~\cite{sloth,lyth,WMAP}. Flat directions which
are lifted only by $d=7,9$ are appropriate for this scenario~\cite{enqvist02}
(see Eq.~(\ref{adpot1}) for $D$-term inflation).

{\it (2)~Fluctuating inflaton coupling:}\\
In a supersymmetric context the inflaton can decay into MSSM degrees of
freedom via vev dependent couplings
\begin{equation}
W \ni \lambda_h I \bar{H} H +
I \frac{q}{M} q q + I \frac{q_c}{M} q_c q_c +
I \frac{h}{M} q q_c,
\end{equation}
where $I$ is the inflaton, $H$ and $\bar{H}$ are the two Higgs
doublets, and $q$ and $q_c$ are (s)quarks and (s)lepton superfields
and their anti-particles. $M$ is the cut-off scale. Fluctuations in
$\delta \rho_{\gamma}/\rho_{\gamma}=-(2/3)\delta\Gamma/\Gamma$
\cite{Dvali,postma,maz}. 
The factor $2/3$ appears due to red-shift of the modes during the decay of the
inflaton whose energy still dominates.
The inflaton decay rate is $\Gamma = 1/ (4\pi) m_I\lambda^2$. The
decay rate fluctuates if either $\lambda$ or $m_I$ is a function of
a fluctuating light field.
\be
\label{rate}
\lambda = \left \{ 
\begin{array}{lll} 
&\lambda_{0} \left( 1 +\frac{\phi}{M}+...\right), 
& \qquad {\rm direct~decay}.
\\
&\lambda_{0} \frac{\phi}{M}, 
& \qquad {\rm indirect~decay}.
\end{array}
\right.
\ee
In the MSSM example, the ``direct'' decay mode corresponds into
inflaton decay into Higgs fields, whereas the ``indirect'' decay mode
corresponds to decay into quarks and anti-squarks.
The fluctuation in the decay rate for these cases is
\be
\label{decay}
\frac{\delta \Gamma}{\Gamma} = \left \{ 
\begin{array}{lll} 
& 2\frac{\delta \phi}{M}, 
& \qquad {\rm direct~decay}.
\\
& 2 \frac{\delta \phi}{\phi}, 
& \qquad {\rm indirect~decay}.
\\
& \frac{\delta \phi}{\phi},
& \qquad {\rm fluctuating~mass}.
\end{array}
\right.  
\ee 
Again fluctuations in MSSM flat direction generates isocurvature fluctuations
which is then transferred to adiabatic fluctuations of the relativistic
species at the time of reheating~\cite{Dvali,Marieke}.

\section{Conclusion}

MSSM along with SUGRA plays multi-faceted role in the early Universe. If
SUSY is the answer for the desert between the electroweak and the 
Planck scale, then MSSM flat directions certainly provide an unique set 
up where questions ranging from density fluctuations to baryon asymmetry 
and cold dark matter generation can be addressed in a most economical way.
This is a great triumph for SUSY and SUGRA that it can not only tame
particle physics domain but can also conquer cosmology.
               

\end{document}